\documentclass[prl,twocolumn,showpacs]{revtex4}
\usepackage[dvips]{graphicx}
\usepackage{amsmath}
\begin{document}

\title{A Molecular Hydrodynamic Theory of 
Supercooled Liquids and Colloidal Suspensions under Shear}
\author{Kunimasa Miyazaki}
\email{miyazaki@fas.harvard.edu}
\author{David R. Reichman}
\email{reichman@fas.harvard.edu}
\affiliation{Department of Chemistry and Chemical Biology,
Harvard University, 12 Oxford Street, Cambridge, MA 02138}

\begin{abstract} 
We extend the conventional mode-coupling theory of supercooled liquids
to systems under stationary shear flow.  Starting from generalized
fluctuating hydrodynamics, a nonlinear equation for the intermediate
scattering function is constructed.  We evaluate the solution
numerically for a model of a two dimensional colloidal suspension and
find that the structural relaxation time decreases as
$\dot{\gamma}^{-\nu}$ with an exponent $\nu \leq 1$, where
$\dot{\gamma}$ is the shear rate.  The results are in qualitative
agreement with recent molecular dynamics simulations.  We discuss the
physical implications of the results.
\end{abstract}
\maketitle

Recently, there has been an explosion of interest in understanding out
of equilibrium properties in supercooled liquids.  In general, the
nonequilibrium behavior of a glassy system is characterized by a
violation of the fluctuation-dissipation theorem and the absence of
time translation invariance.  In the particular case where the system
is subjected to a homogeneous, steady shear flow, time translation
invariance is recovered.  This simpler nonequilibrium situation is of
interest for two reasons.  First, understanding the rheological
properties of complex fluids such as colloidal suspensions and
polymers at a microscopic level is important for the design and
control of new materials.  On a more fundamental level, it has
recently been suggested that for supercooled liquids there are
fundamental connections between the standard thermodynamic control
variables of temperature and density in the equilibrium case and steady
state shear out of equilibrium\cite{liu1998b}.  
A major goal of this work is to
develop a theory that provides an explicit microscopic connection
between the temperature, density and shear rate in a supercooled
liquid.

Dense colloidal suspensions are known to exhibit weak shear thinning
behavior.  Such behavior is predicted for simple liquids as well, but
the effect is too small to observe at temperatures well above the
glass transition.  For supercooled liquids, however, the situation is
different.  Recent numerical simulations have revealed anomalous
rheological behavior in supercooled liquids.  Yamamoto and
Onuki\cite{yamamoto1998c} and Berthier and
Barrat\cite{berthier2002} have simulated supercooled liquids under
strong stationary shear flow and observed a characteristic shear
dependence of the structural relaxation time and the shear viscosity,
$\tau_{\alpha}, \eta \propto \dot{\gamma}^{-\nu}$, where
$\tau_{\alpha}$ is the structural relaxation time, $\eta$ is the shear
viscosity, $\dot{\gamma}$ is the shear rate and the exponent $\nu$ is
empirically found to range between $2/3$ and $1$.  An abstract
schematic approach based on the exactly solvable p-spin spin glass has
been proposed and studied by Berthier, Barrat and
Kurchan\cite{berthier2000b}.  This model predicts $\nu=2/3$ in
agreement with the lower bound found in the simulations of Berthier
and Barrat\cite{berthier2002}.  Since this model is schematic, it
cannot be used to understand in detail the microscopic relationship
between fluid structure and dynamics as a function of thermodynamic
control variables and external driving.

In this letter, we shall generalize the mode-coupling theory developed
to describe the fluctuations in an equilibrium state to that of a
system under a stationary shear flow.  Our starting point is
generalized fluctuating hydrodynamics.  Using several approximations,
we obtain a closed nonlinear equation for the sheared generalization
of the intermediate scattering function.  In this letter, we shall
consider both normal liquids as well as the overdamped Brownian
behavior of a colloidal suspension in the absence of hydrodynamic
interactions.  Numerical results will only be presented for the
Brownian case, but the more general results derived here could be used
to make quantitative contact with recent molecular dynamics
simulations.

Consider the shear flow given by 
\begin{equation}
{\bf v}_{0}({\bf r}) = \mbox{\boldmath$\Gamma$}\cdot{\bf r} = (\dot{\gamma} y, 0, 0), 
\end{equation}
where $(\mbox{\boldmath$\Gamma$})_{\alpha\beta}=\dot{\gamma}\delta_{\alpha x}\delta_{\beta y}$
is the velocity gradient matrix.
The hydrodynamic fluctuations for density $\rho({\bf r}, t)$ and the
velocity field ${\bf v}({\bf r}, t)$ obey the following set of Langevin
equations\cite{kirkpatrick1986c,indrani1995}.  
\begin{equation}
\begin{split}
&
\frac{\partial{\rho}}{\partial t}
= - \nabla\cdot(\rho{\bf v}),
\\
&
\frac{\partial (\rho{\bf v})}{\partial t}
+\nabla\cdot(\rho{\bf v}{\bf v}) 
= 
-\rho \nabla\frac{\delta {\cal F}}{\delta \rho}
-\zeta_{0}({\bf v}-{\bf v}_{0}) + {\bf f}_{R},
\end{split}
\label{eq:hydrodynamcs}
\end{equation}
where $\zeta_{0}$ is the collective friction coefficient for colloidal
particles and ${\bf f}_{R}({\bf r}, t)$ is the random force. 
The $\zeta_{0}$ term is specific for the colloidal case. 
In the case of atomic liquids, the friction term should be replaced by 
a stress term which is proportional to the gradient of velocity field
multiplied by the position dependent shear viscosities.
Both cases, however, lead to the same dynamical behavior at long time
scales.
We neglect the weak ${\bf r}$
dependence of the friction coefficient that could arise from
hydrodynamic interactions between colloidal particles in the Brownian
case.  The first term in the right hand side of the equation for the
momentum is the pressure term and ${\cal F}$ is the total free energy
in a stationary state.  Here we assume that the free energy is well
approximated by that of the equilibrium form and is given by a
well-known expression;
\begin{equation}
\begin{split}
\beta{\cal F} 
\simeq 
&
\int\!\!\mbox{d}{\bf r}~ \rho({\bf r})\left\{ \ln\rho({\bf r}) -1 \right\}
\\
&
-\frac{1}{2}\int\!\!\mbox{d}{\bf r}_1\int\!\!\mbox{d}{\bf r}_2~
\delta\rho({\bf r}_1)c(r_{12})\delta\rho({\bf r}_2),
\end{split}
\end{equation}
where $\beta=1/k_{\mbox{\scriptsize B}} T$ and $c(r)$ is the direct correlation function.
Under shear, it is expected that $c(r)$ will be distorted and should
be replaced by a nonequilibrium, steady state form
$c_{\mbox{\scriptsize noneq}}({\bf r})$, which is an anisotropic function of
${\bf r}$.  The effect of shear on the static correlation functions
has been studied\cite{onuki1997,ronis1986} and is found that the
distortion is characterized by the P\'{e}clet number Pe$=\dot{\gamma}
\sigma^2/D_0$, where $\sigma$ is the diameter of the particle and
$D_0=k_{\mbox{\scriptsize B}} T/\zeta_0$ 
is the diffusion constant.  
Thus, if the
P\'{e}clet number is small, the above assumption is valid.  If necessary,
the full anisotropic steady state structure may be used.
By linearizing eq.(\ref{eq:hydrodynamcs}) around the stationary state as
$\rho = \rho_{0}+\delta \rho$ and ${\bf v} = {\bf v}_{0} + \delta{\bf v}$, where
$\rho_{0}$ is the average density, we obtain the following equations,
\begin{equation}
\begin{split}
&
\left(
\frac
{\partial~}{\partial t}- {\bf k}\cdot\mbox{\boldmath$\Gamma$}\cdot
\frac{\partial ~}{\partial {\bf k}}
\right)
\delta\rho_{{\bf k}}(t) = -ik J_{{\bf k}}(t),
\\
&
\left(
\frac{\partial ~}{\partial t}
- {\bf k}\cdot\mbox{\boldmath$\Gamma$}\cdot
 \frac{\partial ~}{\partial {\bf k}}
+ \hat{\bf k}\cdot\mbox{\boldmath$\Gamma$}\cdot\hat{\bf k}
\right)
J_{{\bf k}}(t)
\\
&
= 
-\frac{ik}{m\beta S(k)}\delta\rho_{{\bf k}}(t)
\\
&
-\frac{1}{m\beta}
\int_{{\bf q}}i\hat{\bf k}\cdot{\bf q} c(q)\delta\rho_{{\bf k}-{\bf q}}(t)\delta\rho_{{\bf q}}(t)
-\frac{\zeta_{0}}{m}J_{{\bf k}}(t) + {\bf f}_{R{\bf k}}(t),
\end{split}
\label{eq:nonlinear}
\end{equation}
where 
$c(q)$ is the Fourier transform
of $c(r)$, 
$\hat{\bf k} \equiv {\bf k}/|{\bf k}|$,
$J_{{\bf k}}(t)=\rho_{0}\hat{\bf k}\cdot\delta{\bf v}_{{\bf k}}(t)$ is the
longitudinal momentum fluctuation, 
and 
$\int_{{\bf q}}\equiv \int\mbox{d}{\bf q}/(2\pi)^d$ for a 
$d$-dimensional system.
Note that our approximate equation does not contain
coupling to transverse momentum fluctuations even in the presence of
shear.

In order to construct equations for the appropriate correlations
from the above expressions, an approximate symmetry is necessary.  In
the presence of shear, translational invariance is violated.  In other
words, correlations of arbitrary fluctuations, $f({\bf r},t)$ and $g({\bf r},
t)$, do not satisfy $\langle f({\bf r},t)g({\bf r}', 0)\rangle \neq \langle
f({\bf r}-{\bf r}',t)g({\bf 0},0)\rangle$.  Instead, we shall assume that the
following symmetry is valid\cite{onuki1997},
\begin{equation}
\langle f({\bf r}, t)g({\bf r}', 0) \rangle
= 
\langle f({\bf r}-{\bf r}'(t), t)g({\bf 0}, 0) \rangle,
\label{eq:transinv}
\end{equation}
where we defined the time-dependent position vector by 
${\bf r}(t) \equiv \exp[\mbox{\boldmath$\Gamma$} t]\cdot{\bf r} 
= {\bf r} + \dot{\gamma} t y \hat{\bf e}_{x}$, where 
$\hat{\bf e}_{x}$ is an unit vector oriented along the $x$-axis.
In wave vector space, this is expressed as 
\begin{equation}
\begin{split}
\langle f_{\bf k}(t)g^{*}_{{\bf k}'}(0) \rangle
&
= 
\langle f_{{\bf k}}(t)g^{*}_{{\bf k}(t)}(0) \rangle
\times \delta_{{\bf k}(t),{\bf k}'}
\\
&
= 
\langle f_{{\bf k}'(-t)}(t)g^{*}_{{\bf k}'}(0) \rangle
 \times \delta_{{\bf k},{\bf k}'(-t)},
\end{split}
\label{eq:inv}
\end{equation}
where ${\bf k}(t) = \exp[{}^{t}\mbox{\boldmath$\Gamma$} t]\cdot{\bf k} = {\bf k} + \dot{\gamma} t
k_{x}\hat{\bf e}_{y}$, 
where ${}^{t}\mbox{\boldmath$\Gamma$}$ denotes the transpose of
$\mbox{\boldmath$\Gamma$}$ and $\delta_{{\bf k},{\bf k}'}\equiv
(2\pi)^{d}V^{-1}\delta({\bf k}-{\bf k}')$ for a system of 
volume $V$.
 
Eq.(\ref{eq:transinv}) states that the fluctuations satisfy
translational invariance in a reference frame flowing with the shear
contours.  This approximation holds for long wavelengths where the
direct interactions between particles are not important.  On the other
hand, for correlations between particles separated by molecular length
scales, this is not generally true.  The validity of this
approximation for molecular length scales should be systematically
examined in the future.

Using this approximation, it is straightforward to construct the
mode-coupling equations for the appropriate correlation functions. 
We shall derive the equation for the intermediate scattering function 
defined by 
\begin{equation}
F({\bf k}, t) \equiv \frac{1}{N}
\langle \delta\rho_{{\bf k}(-t)}(t)\delta\rho^{*}_{{\bf k}}(0) \rangle,
\label{eq:Fkt}
\end{equation}
where $N$ is the total number of the particles in the system.
Note that 
the wave vector in $\delta\rho_{{\bf k}}(t)$ is now replaced  
by a time-dependent one ${\bf k}(-t)$.

Eq.(\ref{eq:nonlinear}) has a nonlinear term 
\begin{equation}
R_{{\bf k}}(t)
= -\frac{1}{m\beta}\int_{{\bf q}}i(\hat{\bf k}\cdot{\bf q})c({\bf q})
  \delta\rho_{{\bf k}-{\bf q}}(t)\delta\rho_{{\bf q}}(t).
\label{eq:Rk}
\end{equation}
This term can be renormalized 
with the definition of a generalized friction coefficient 
following the standard procedure of derivation of the 
mode-coupling equations\cite{kirkpatrick1986c,zwanzig2001}. 
To lowest order in the fluctuations, we have
\begin{equation}
\begin{split}
&
\left(
\frac{\partial ~}{\partial t}
- {\bf k}\cdot\mbox{\boldmath$\Gamma$}\cdot\frac{\partial~}{\partial {\bf k}}
+ \hat{\bf k}\cdot\mbox{\boldmath$\Gamma$}\cdot\hat{\bf k}
\right)
J_{\bf k}(t)
= 
-\frac{ik}{m\beta S(k)}\delta\rho_{{\bf k}}(t)
\\
&
-\frac{1}{m}\int_{-\infty}^{t}\!\!\mbox{d} t'~
\int\!\!\mbox{d}{{\bf k}'}~
\zeta({\bf k}, {\bf k}', t-t')J_{{\bf k}'}(t')
+ {\bf f}'_{R{\bf k}}(t),
\end{split}
\label{eq:MCT}
\end{equation}
where $\zeta({\bf k}, {\bf k}', t)$ is the generalized friction coefficient and 
${\bf f}'_{R{\bf k}}(t)$ is a corresponding random force. 
$\zeta({\bf k}, {\bf k}', t)$ is given by the sum of the bare friction
coefficient and the mode-coupling term as
\begin{equation}
\zeta({\bf k}, {\bf k}', t)
= \zeta_{0}\times 2\delta(t)+\delta\zeta({\bf k}, {\bf k}', t)
\end{equation}
with the mode-coupling contribution given by 
\begin{equation}
\delta\zeta({\bf k}, {\bf k}', t)
= 
\frac{m^2\beta}{N}\langle R_{\bf k}(t)R^{*}_{{\bf k}'}(0) \rangle.
\end{equation}
Substituting eq.(\ref{eq:Rk}) into the above expression, we obtain
\begin{equation}
\begin{split}
&
\delta\zeta({\bf k},{\bf k}' t)
= 
\frac{1}{\beta N}\int_{{\bf q}}\int_{{\bf q}'}
\hat{\bf k}\cdot{\bf q} c({\bf q})\hat{\bf k}'\cdot{\bf q}'c({\bf q}')
\\
&
\hspace*{0.5cm}
\times
\langle 
\delta\rho_{{\bf k}-{\bf q}}(t)
\delta\rho_{{\bf q}}(t)
\delta\rho^{*}_{{\bf k}'-{\bf q}'}(0)
\delta\rho^{*}_{{\bf q}'}(0) 
\rangle.
\end{split}
\label{eq:deltaG}
\end{equation}
This involves a four point correlation function. 
Using the Gaussian approximation, 
this can be decomposed into a product of 
two-point correlation functions as 
\begin{equation}
\begin{split}
&
\langle 
\delta\rho_{{\bf k}-{\bf q}}(t)\delta\rho_{{\bf q}}(t)
\delta\rho^{*}_{{\bf k}'-{\bf q}'}(0)\delta\rho^{*}_{{\bf q}'}(0) 
\rangle
\\
&
\simeq
N^2
F({\bf k}(t)-{\bf q}(t), t)F({\bf q}(t),t)
\\
&
\hspace*{0.5cm}
\times
\left\{
\delta_{{\bf k}(t),{\bf k}'}\delta_{{\bf q}(t),{\bf q}'}
+
\delta_{{\bf k}(t),{\bf k}'}\delta_{{\bf q}(t),{\bf k}'-{\bf q}'}
\right\},
\end{split}
\label{eq:n4}
\end{equation}
where use has been made of 
the translational invariance, eq.(\ref{eq:inv}).

Substituting eq.(\ref{eq:n4}) back to 
eq.(\ref{eq:deltaG}), we obtain
\begin{equation}
\delta\zeta({\bf k},{\bf k}',t)
= \delta\zeta({\bf k},t)\delta_{{\bf k}(t),{\bf k}'}
\end{equation}
with
\begin{equation}
\begin{split}
&
\delta\zeta({\bf k}, t)
= 
\frac{\rho_{0}}{\beta}
\int_{{\bf q}}
\left\{
\hat{\bf k}\cdot{\bf q} c({\bf q})+\hat{\bf k}\cdot\left({\bf k}-{\bf q} \right)c({\bf k}-{\bf q})
\right\}
\\
&
\hspace*{1.0cm}
\times
 \hat{\bf k}(t)\cdot{\bf q}(t) c({\bf q}(t)) 
F({\bf k}(t)-{\bf q}(t), t)F({\bf q}(t), t) 
\\
&
= 
\frac{\rho_{0}}{2\beta}
\int_{{\bf q}}
{\cal V}({\bf k}, {\bf q}){\cal V}({\bf k}(t), {\bf q}(t))
F({\bf k}(t)-{\bf q}(t), t)F({\bf q}(t), t), 
\end{split}
\end{equation}
where ${\cal V}({\bf k}, {\bf q})$ is the vertex function given by 
\begin{equation}
{\cal V}({\bf k}, {\bf q})
=
\hat{\bf k}\cdot
\left\{
{\bf q} c({\bf q}) + \left({\bf k}-{\bf q} \right)c({\bf k}-{\bf q}) 
\right\}.
\end{equation}
From these results and eq.(\ref{eq:MCT}), 
the equation for the correlation function,
\[
C({\bf k}, t) 
\equiv 
\frac{1}{N}
\langle J_{{\bf k}(-t)}(t) n^{*}_{{\bf k}}(0) \rangle
\]
is given by 
\begin{equation}
\begin{split}
&
\frac{\mbox{d} C({\bf k}, t) }{\mbox{d} t}
\\
&
= 
-\hat{\bf k}(-t)\cdot\mbox{\boldmath$\Gamma$}\cdot\hat{\bf k}(-t)C({\bf k}, t)
- \frac{ik(-t)}{m\beta S(k(-t))}F({\bf k}, t)
\\
&
-\frac{1}{m}\int_{0}^{t}\!\!\mbox{d} t'~
\delta \zeta({\bf k}(-t),t-t')C({\bf k}, t').
\end{split}
\label{eq:C}
\end{equation}
Note that in the above equation, the differential operator 
${\bf k}\cdot\mbox{\boldmath$\Gamma$}\cdot\partial/\partial{\bf k}$ disappears because
\begin{equation}
\frac{\mbox{d} C({\bf k}, t) }{\mbox{d} t}
= \frac{\partial C({\bf k}, t) }{\partial t}
 -{\bf k}(-t)\cdot\mbox{\boldmath$\Gamma$}
  \cdot\frac{\partial~}{\partial{\bf k}(-t)}C({\bf k}, t).
\label{eq:}
\end{equation}
Likewise, the continuity equation (the first term in
eq.(\ref{eq:nonlinear})) can be written as
\begin{equation}
\begin{split}
\frac{\mbox{d}F({\bf k}, t) }{\mbox{d}t}
&
= 
-ik(-t) C({\bf k}, t).
\end{split}
\label{eq:cont}
\end{equation}
This equation together with eq.(\ref{eq:C})
comprises the closed set of the 
mode-coupling equations for $F({\bf k},t)$ and 
$C({\bf k}, t)$ under shear. 

For colloidal suspensions the relaxation time of the momentum
fluctuations is of the order of $\tau_{m}= m/\zeta_{0}$ and is much
shorter than the relaxation time for density fluctuations which is of
the order of or longer than $\tau_{d}=\sigma^2/D_{0}$.  For the time
scale of interest, $\hat{\bf k}(-t)\cdot\mbox{\boldmath$\Gamma$}
\cdot\hat{\bf k}(-t)$ as well as the
inertial term can be neglected in the equation for the momentum
fluctuations since $\dot{\gamma} \tau_{m} \ll 1$ in realistic situations.  
Thus, the equation for
the momentum fluctuations may be written as
\begin{equation}
\begin{split}
0
= 
&
-\frac{ik(-t)}{\beta S(k(-t))}F({\bf k}, t)
-\zeta_{0}C({\bf k}, t)
\\
&
-\int_{0}^{t}\!\!\mbox{d} t'~
\delta\zeta({\bf k}(-t),t-t')C({\bf k}, t').
\end{split}
\end{equation}
Substituting this back into eq.(\ref{eq:cont}), 
we arrive at
\begin{equation}
\begin{split}
\frac{\mbox{d} F({\bf k}, t) }{\mbox{d}t}
= 
&
-\frac{D_{0}k(-t)^2}{S(k(-t))}F({\bf k}, t)
\\
&
-\int_{0}^{t}\!\!\mbox{d} t'~
  M({\bf k}(-t),t-t')
  \frac{\mbox{d} F({\bf k}, t') }{\mbox{d}t'},
\end{split}
\label{smolchowski}
\end{equation}
where 
\begin{equation}
\begin{split}
M({\bf k},t)
= 
&
\frac{\rho_{0}D_{0}}{2}
\frac{k}{k(t)}
\int_{{\bf q}}
{\cal V}({\bf k}, {\bf q}){\cal V}({\bf k}(t), {\bf q}(t))
\\
&
\times
F({\bf k}(t)-{\bf q}(t), t)F({\bf q}(t), t).
\end{split}
\label{eq:Mkt}
\end{equation}
Eqs.(\ref{smolchowski}) and (\ref{eq:Mkt}) 
are the major result of this letter. 
In the absence of the shear, they reduce to the conventional 
mode-coupling equations\cite{gotze1992}.  

In order to study the shear thinning effect in the supercooled state,
we shall solve eqs.(\ref{smolchowski}) and (\ref{eq:Mkt}) numerically.
Solving this equation is more difficult than solving the corresponding
equation in the equilibrium case because the wave vectors are
distorted by shear and the system is not isotropic.  For simplicity,
we shall consider a hypothetical two-dimensional colloidal suspension
which is simple to handle numerically but still undergoes an
ergodic-nonergodic transition below a certain density.  The shear flow
occurs in the $x$ direction.  We have chosen the following form of the
static structure factor $S(k)$:
\begin{equation}
S(k) = S_{\mbox{\scriptsize PY}}(k+k_0, \alpha\rho_{0})f(k-k_c), 
\end{equation} 
where $S_{\mbox{\scriptsize PY}}(k, \rho)$ is the static structure
factor for a hard-sphere system at the density $\rho$
obtained from the Percus-Yevick closure, 
$k_{0}$ and $\alpha$ are parameters which were chosen in such a way that
$S(k)$ is short-ranged and has broader peak.
$f(k-k_c)$ is a cut-off function which makes $S(k)$ approach unity
smoothly for wave vectors larger than the cut-off $k_c$.
The choice of $S(k)$ mimics the
shape of $S(k)$ of 
real systems although it does not satisfy 
sum-rule restrictions.
In our calculation, we chose $k_0=4.0$, $\alpha=32$ and $k_{c}=4.0$.
For this system, the ergodic-nonergodic transition occurs around
a ``density'' $\rho_{c}\sigma^2\simeq 1.2\times 10^{-2}$ in the absence
of shear.
In Figure 1, we show the behavior of $F({\bf k}, t)$ 
for $\rho\sigma^2=1.15\times 10^{-2}$, 
slightly below $\rho_{c}$. 
The wave vector is ${\bf k} = (0, 2\sigma)$.  Since $k_{x}=0$, the
expression for $F({\bf k}, t)$ is equivalent to that in the quiescent
state, $F_{\mbox{\scriptsize eq}}({\bf k}, t) \equiv N^{-1} \langle
\delta\rho_{{\bf k}}(t)\delta\rho^{*}_{{\bf k}}(0) \rangle$.  Thus,
there is no direct effect from the convection term but, due to the
nonlinear coupling through the mode-coupling term $M({\bf k}, t)$, a
strong shear dependence of the relaxation time can be seen.  The
dependence of the structural relaxation time
$\tau_{\alpha}(\dot{\gamma})$ on the shear rate $\dot{\gamma}$ is
estimated from the value where $F({\bf k}, \tau_{\alpha}) =
\mbox{\large e}^{-1}$.  For the particular case of
$\rho\sigma^2=1.15\times 10^{-2}$ we find the power law
$\tau_{\alpha}\simeq \dot{\gamma}^{-\nu}$ with $\nu\simeq 0.8$ for
Pe$\geq 10^{-3}$ and $\tau_{\alpha}$ saturates to the equilibrium
value at Pe$\leq 10^{-3}$.  This is similar to the behavior reported
in recent molecular dynamics simulations.  For values of $\rho\sigma^2
> 1.2\times 10^{-2}$ (not shown), we find that the exponent $\nu$
saturates at a higher value, in agreement with the simulations of
Berthier and Barrat\cite{berthier2002}.

The physical picture that emerges from the molecular hydrodynamic
theory developed here is simple.  The shear flow perturbs and
randomizes the coupling between different modes.  Physically, this
perturbation dissipates the cage that transiently immobilizes
particles.  Mathematically, this is reflected through the time
dependence of the vertex, which vanishes as $t \rightarrow \infty$.
This simple picture illustrates the essence of the mode-coupling
approach to the shear thinning effect in simple supercooled systems.
Note that even for fluctuations orthogonal to the direction of flow,
thinning occurs due to the coupling of fluctuations in all directions.
In this sense, the picture of cage breakup in a supercooled liquid due
to external flow is quite different from that of dynamic critical
phenomena under shear, in which the faster relaxation occurs solely
because the fluctuations are stretched out by the shear flow and
pushed to larger wave vectors where faster relaxation occurs.
 
In this letter, we have derived an approximate mode-coupling theory
for a supercooled liquid under steady shear flow.  The most important
assumption is the use of approximate translational invariance,
eq.(\ref{eq:inv}).  This allows one to derive a nonlinear
integro-differential equation for $F({\bf k}, t)$ similar to the one
for the equilibrium state.  The numerical analysis for a hypothetical
two-dimensional colloidal suspension has been carried out and a
typical behavior of $F({\bf k}, t)$ was shown to be consistent with
recent simulations.  The relaxation time is found to have the strong
shear dependence.  More systematic and thorough analysis of the
numerical solutions of the mode-coupling equations are left for
future work\footnote{
As this manuscript was being written for publication, we became aware
of a parallel effort by Fuchs and Cates
[cond-mat/0204628;cond-mat/0207530].  While the omission of several
details in their calculation make a detailed comparison between the
two approaches impossible at this time, there are many similarities in
the final expressions, and the physical picture is also quite similar.
Differences do exist in the expressions, which most likely lead to
different estimates of the exponent $\nu$.  It will be most useful in
the future to make a detailed comparison of the two theories since the
differences in approach (fluctuating hydrodynamics verses the
projection operator formalism) and the final results may lead to a
better understanding of the limitations of several assumptions used to
here and by Fuchs and Cates to describe how an external flow
disrupts jamming.}.

\vspace*{0.5cm}

The authors acknowledge support from NSF grant \#0134969.  The authors
would like to express their gratitude to Dr. Ryoichi Yamamoto for
suggesting this problem during his stay at Harvard, and for useful
discussions.

\newpage

\begin{center}
{\bf\large FIGURE CAPTIONS}
\end{center}

\vspace*{1.0cm}

\begin{figure}[ht]
\caption{Normalized $F({\bf k}, t)$ for ${\bf k}=(0, 2\sigma)$ for various 
shear rates $\dot{\gamma}$.
The ``density'' is $\rho\sigma^2 = 1.15\times 10^{-2}$. 
From the right to the left, 
$\mbox{Pe} =\dot{\gamma}\sigma^2/D_{0} = 0$, $10^{-4}$, 
$10^{-3}$, $10^{-2}$, $10^{-1}$, and $1$.
The results for Pe$=0$ and $10^{-4}$ are almost indistinguishable.
The time $t$ is scaled by $\sigma^2/D_{0}$. 
} 
\end{figure}

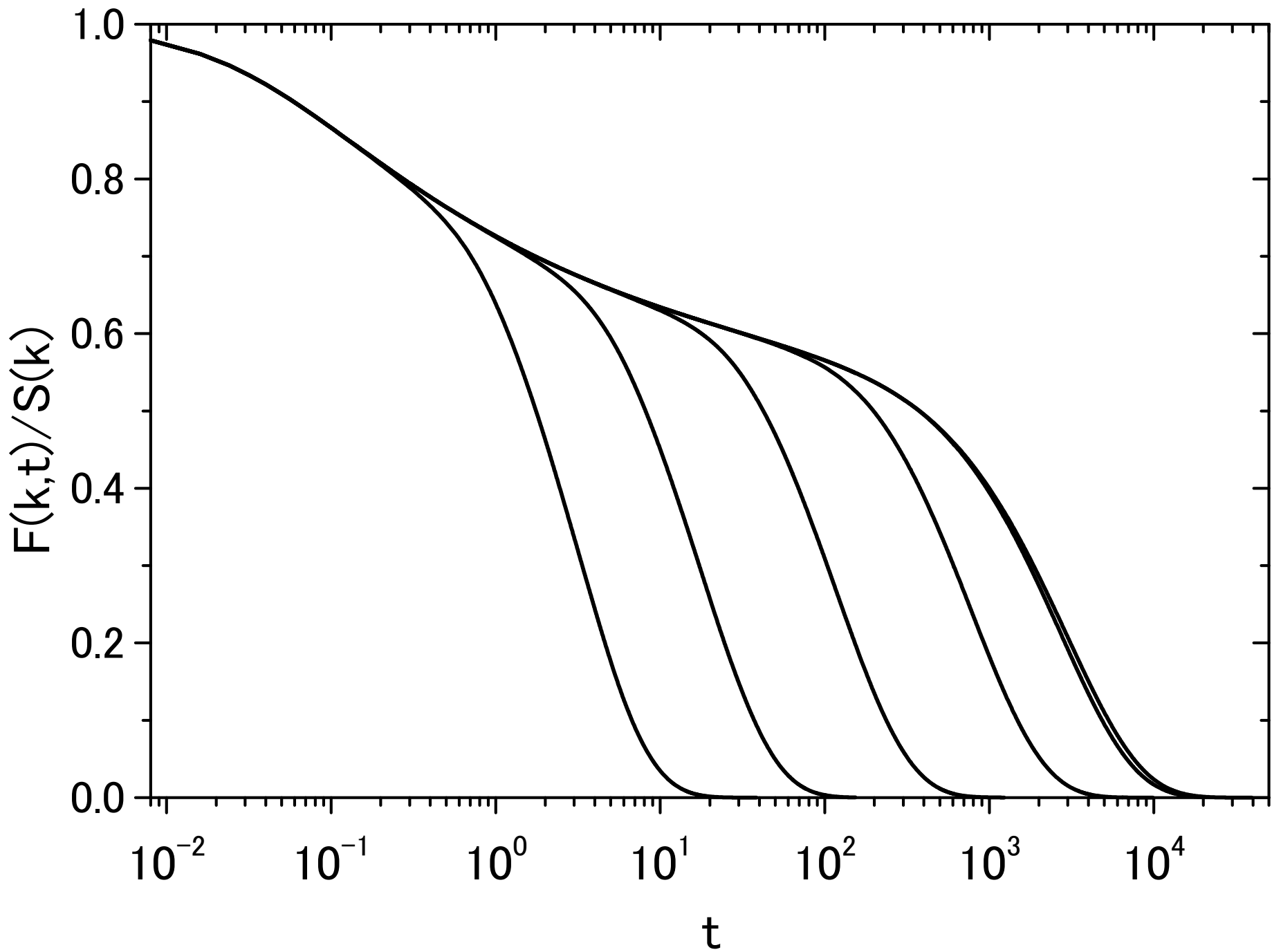
\end{document}